**An *Ab Initio* Study of the Adsorption and Dissociation of Molecular Oxygen**

**on the (0001) Surface of Double Hexagonal Close Packed Americium**

Pratik P. Dholabhai, Raymond Atta-Fynn and Asok K. Ray*

Physics Department, University of Texas at Arlington,

Arlington, Texas 76019

*akr@uta.edu.

**Abstract**

In our continuing attempts to understand theoretically various surface properties such as corrosion and potential catalytic activity of actinide surfaces in the presence of environmental gases, we report here the *first ab initio study of molecular adsorption* on the double hexagonal packed (dhcp) americium (0001) surface. Specifically, molecular oxygen adsorption on the (0001) surface of dhcp americium (Am) has been studied in detail within the framework of density functional theory using a full-potential all-electron linearized augmented plane wave plus local orbitals method (FP-L/APW+lo). Dissociative adsorption is found to be energetically more favorable compared to molecular adsorption. Chemisorption energies were optimized with respect to the distance of the adsorbates from the surface for three approach positions at three adsorption sites, namely t1 (one-fold top), b2 (two-fold bridge), and h3 (three-fold hollow) sites. Chemisorption energies were computed at the scalar-relativistic-no-spin-orbit-coupling (SR-NSOC) and at the fully-relativistic-with-spin-orbit-coupling (FR-SOC) levels of theory. The most stable configuration corresponds to a horizontal approach molecular dissociation with the oxygen atoms occupying neighboring h3 sites, with chemisorption energies at the NSOC and SOC theoretical levels being 9.395 eV and 9.886 eV, respectively. The corresponding distances of the oxygen molecule from the surface and oxygen – oxygen distance were found to be 0.953 Å and 3.731 Å, respectively. Overall our calculations indicate that chemisorption energies in cases with SOC are slightly more stable than the cases with NSOC in the 0.089-0.493 eV range. The work functions and net magnetic moments respectively increased and decreased in all cases compared with the corresponding quantities of the bare dhcp Am (0001) surface. The adsorbate-substrate interactions have been analyzed in detail using the partial charges inside the muffin-tin spheres, difference charge density distributions, and the local density of states. The effects, if any, of chemisorption on the Am $5f$ electron localization-delocalization characteristics in the vicinity of the Fermi level are also discussed.

**Keywords:** Density Functional Theory; Molecular and Dissociative Adsorption; Double Hexagonal Close Packed Americium; Molecular Oxygen.

**PACS:** 71.20.-b; 68.35.-p; 71.27.+a; 68.43.-h

**1 Introduction**

Surface chemistry and physics have been and continue to be very active fields of research because of the obvious scientific and technological implications and consequent importance of such research [1]. One of the *many* motivations for this burgeoning effort has been the desire to understand surface corrosion, metallurgy and catalytic activity in order to address environmental concerns. In particular, such efforts are important for the actinides, for which experimental work is relatively difficult to perform due to material problems and toxicity. *Some* of the actinides are among the most complex of long-lived elements, and in the solid state, display most unusual behaviors of any series in the periodic table. Plutonium (Pu), for instance, has a relatively low melting temperature, large anisotropic thermal expansion coefficients, very low symmetry crystal structures at low temperatures, and many solid-to-solid phase transitions as a function of temperature. On the other hand, americium (Am) exhibits solid-to-solid phase transitions as a function of pressure, with the high pressure phases having low symmetry crystal structures. Radioactive and highly electropositive, the actinides are characterized by the gradual filling of the 5*f* electron shell with the degree of localization increasing with the atomic number Z along the last series of the periodic table. The light actinides, from thorium to plutonium, are believed to have itinerant 5*f* electrons participating in metallic bonding and the heavier actinides continuing from americium have localized 5*f* electrons. The open shell of the 5*f* electrons determines the magnetic and solid state properties of the actinides and their compounds. However, these properties of the actinides, particularly the transuranium actinides, are *still not* clearly understood. This stems primarily from the inherent difficulty in understanding the delocalization-to-localization transition in the behavior of the 5*f* electrons as one proceeds from Pu to Am and the dramatic increase in atomic volume. The actinides are also characterized by the increasing prominence of relativistic effects and their study can, in fact, give us an in depth understanding of the role of relativity throughout the periodic table [2-8].

Among the transuranium actinides, the unique electronic properties of the manmade Americium (Am) metal, which was first successfully synthesized and isolated at the wartime Metallurgical Laboratory [9], have received increased interests recently, from both scientific and technological points of view. As stated above, Am occupies a central position in the actinide series in our understanding of the behavior of the *5f* electrons. It is widely believed, as mentioned above, that the properties and the behavior of the *5f* electrons change dramatically starting from somewhere between Pu and Am. As a result, a large number of experimental and theoretical works have been done in recent years to gain insight into the structural and electronic properties of Am. We have listed only *some* of the representative works in Refs. 10-27. Critical and important questions related to Am are phase transitions with increasing pressure, localization/delocalization behavior of the *5f* electrons, and *possible* magnetism of the different phases. We have discussed these issues and the relevant literature, in detail, in our previous work on the quantum size effects in fcc and dhcp Am [28]. In particular, the anti-ferromagnetic state with spin-orbit coupling was found to be the ground state of dhcp Am with the *5f* electrons primarily localized and the surface energy and work function of the of the dhcp Am(0001) surface were predicted to be 0.84 J/m$^2$ and 2.90eV. Quite recently, we studied the adsorption of atomic hydrogen and oxygen of the (0001) surface of dhcp Am [29]. As a continuation of our systematic density functional studies of environmental gas interactions with actinide surfaces [30], we report, in this work, the adsorption and possible dissociation of molecular oxygen on the (0001) surface of dhcp Am. Our aim is to identify chemically stable binding sites, probe the adsorbate-induced changes in the surface electronic and magnetic structures, and the corresponding effects of such changes, if any, on the Am *5f* electron localization-delocalization. To the best of our knowledge, *no* such study exists in the literature, though, as mentioned in our previous works, an effective way to probe the actinides' *5f* electron properties and their roles in chemical bonding is a study of their bare surface properties and atomic and molecular adsorptions on them.

## 2 Computational Method

All calculations have been performed within the generalized gradient approximation to density functional theory [31] using the Perdew-Burke-Ernzerhof (PBE) formulation for the exchange-correlation functional [32]. The Kohn-Sham equations were solved using the full-potential linear augmented plane wave plus local basis (FP-LAPW+lo) method as implemented in the WIEN2k code [33]. This method makes no shape approximation to the potential or the electron density.

Within the FP-LAPW+lo method, the unit cell is divided into non-overlapping muffin-tin spheres and an interstitial region. Inside the muffin-tin sphere of radius $R_{MT}$, the wave functions are expanded using radial functions (solutions to the radial Schrödinger equation) times spherical harmonics with angular momenta up to $l_{max}^{wf}$ =10. Non-spherical contributions to the electron density and potential inside the muffin tin spheres were considered up to $l_{max}^{pot}$ =6. APW+lo basis were used to describe *s, p, d,* and *f* (*l* = 0, 1, 2, 3) states and LAPW basis were used for all higher angular momentum states in the expansion of the wave functions. Additional local orbitals (LO) were added to the 2*s* semi-core states of O and the 6*s*, 6*p* semi-core states of Am to improve their description.

The radii of the muffin-tin spheres used were $R_{MT}$(O) = 1.1 Bohr and $R_{MT}$(Am) = 2.2 Bohr. The truncation of the modulus of the reciprocal lattice vector used for the expansion of the wave function in the interstitial region $K_{MAX}$, was set to $R_{MT}K_{MAX}$ = 8.5 for the clean slab and $R_{MT} \times K_{MAX}$ = 4.25 for the slab-with-molecule, where $R_{MT}$ denotes the smallest muffin-tin radius, that is, $R_{MT}$ = 2.2 Bohrs for the bare slab and $R_{MT}$ = 1.1 Bohrs for the slab-with-molecule (this ensures that $K_{MAX}$ is the same each case).

In the WIEN2k code, core states are treated at the fully relativistic level. Semi-core and valence states are treated at either the scalar relativistic level, i.e., no spin-orbit coupling (NSOC) or at the fully relativistic level, i.e., spin-orbit coupling (SOC) included. Spin-orbit interactions for semi-core and valence states are incorporated via a second variational procedure [34] using the scalar relativistic eigenstates as basis, where all eigenstates with energies below a cutoff energy of 4.5 Ry were included, with the so-called $p_{1/2}$ extension [35], which accounts for the finite character of the wave function at the nucleus for the $p_{1/2}$ state. The spin quantization axis for the magnetic SOC calculations was along the [001] direction. We considered both NSOC and SOC cases to investigate spin-orbit coupling effects on chemisorption energies. We have used this method extensively in *some* of our previous works on actinide surfaces [28-30].

The dhcp-Am (0001) surface was modeled by a supercell consisting of a periodic 6-layer slab with a (2×2) surface unit cell and a vacuum 30 a.u. thick (Fig.1). In accordance with our previous findings [29], we have used an AFM configuration for the slab which consists of alternating ferromagnetic layers of up- and down-spin atoms along the *c*-axis. The relaxation of the surface was carried out in two steps: first, bulk dhcp Am was optimized followed by surface optimization. The atomic volume of bulk dhcp Am was expressed in terms a single lattice constant by constraining the ratio *c*/*a* to match experimental value. More precisely, the ratio *c*/*a* was set to 3.2 (experimental ratio) and the volume V was expressed in terms of only *a*. Then the total energy *E* (for an AFM configuration) was computed for several variations of *V*. The energy versus volume E-V fit via Murnaghan's equation of state [36] generated an equilibrium volume $V_o$ = 208.6 (a.u.)$^3$ and B = 25.4 GPa. The equilibrium volume $V_o$ corresponded to *a*=6.702 a.u. The experimental values are 198.4 (a.u.)$^3$ or 197.4 (a.u.)$^3$ corresponding to *a*=6.56 a.u. and 29.7 GPa [11.15, 37].

Using the optimized lattice constants, that is, $a$=6.702 a.u. and c = 3.2$a$, a 2x2 hexagonal surface unit cell for (0001) orientation is constructed. Then the surface unit cell is used to build the slab with 6 atomic layers and 30 a.u. vacuum. Furthermore, the slab was built to have inversion symmetry for computational efficiency. The interlayer spacing between the surface unit cells in the slab above corresponded to a bulk spacing of $d_0$= $c/4$. To relax the slab, the two central layers were fixed at the bulk positions and the 2 outermost layers (this is the same from both sides of the central slab because of inversion symmetry) were allowed to relax. Specifically, the relaxation was performed by minimizing the total energy with respect to $d_{12}$, the separation between the central and subsurface layers, and $d_{23}$, the separation distance between the subsurface and surface layers. Variations of -4%, -2%, 0%, 2%, 4%, measured in terms of the bulk interlayer spacing $d_0$ for $d_{12}$ and $d_{23}$ was used, resulting in a 5x5 grid for the energy computation. The amount of relaxations (Fig.2) obtained were $\Delta d_{12}/d_0$ = 0% and $\Delta d_{23}/d_0$ = 2%, with the reduction in the total energy of the slab being 2.19 mRy. The small relaxations and reduction in the total energy indicate the general stability of the surface.

Integrations in the Brillouin zone (BZ) have been performed using the special k-points sampling method with a temperature broadening of the Fermi surface by the Fermi distribution, where a broadening parameter of $K_B T$ = 0.005 Ry has been used. The temperature broadening scheme avoids the instability from level crossings in the vicinity of the Fermi surface in metallic systems and also reduces the number of k-points necessary to calculate the total energy of metallic systems [38]. For the present work, a 6×6×1 k-mesh density (18 k-points in the irreducible part of the BZ) was deemed to be sufficient. Self-consistency was achieved when the total energy variation from iteration to iteration was 0.01 mRy or lower.

To study adsorption on the *relaxed* Am surface, the admolecule, corresponding to a surface coverage of $\Theta$ = 1/4 of a monolayer (ML), was allowed to approach the surface from both sides to preserve inversion symmetry. Three high symmetry adsorption sites were considered: (i) one-fold top site t1 (admolecule is directly on top of a Am atom) (ii) two-fold bridge site b2 (admolecule is placed in the middle of two nearest neighbor Am atoms); and (iii) three-fold hollow hcp site h3 (admolecule sees a Am atom located on the layer directly below the surface layer). For each of these three adsorption sites, three approaches of the $O_2$ molecule towards to the surface were considered: (a) approach vertical to the surface (Vert approach); (b) approach parallel to a lattice vector (Hor1 approach); (c) approach perpendicular to a lattice vector (Hor2 approach). It is obvious that for both the horizontal approaches the atoms of the oxygen molecule are at the same distance from the americium surface, whereas for the vertical approach one oxygen atom is closer to the surface than the other. With these choices of the surface and the ad-molecule, the oxygen-oxygen interaction between cell repetitions is not expected to be significant.

For geometry optimizations, the distances of the oxygen atoms from the surface $(R_d)$ and the distance between the oxygen atoms $(R_o)$ were simultaneously optimized. The chemisorption energy $E_C$ is given by:

$$E_C(R_d, R_o) = 1/2 \, [E(Am) + 2E(O_2) - E(Am+O_2)],$$

where $E(Am)$ is the total energy of the bare Am slab, $E(O_2)$ is the total energy of the oxygen molecule at the optimized bond length of 1.217Å, and $E(Am+O_2)$ is the total energy of the slab-with-molecule. Thus a positive value of $E_C$ implies chemisorption and a negative value implies otherwise. To calculate the total energy of the ad-molecule, the molecule was fully relaxed in a large box of side 30 Bohr and at the Γ k-point, with all other computational parameters remaining the same.

We wish to note here that no additional relaxations were taken into account primarily because of the all-electron nature of the calculations and the consequent extreme computational effort required. Also, from our recent findings [29], the difference in chemisorption energies between the system where the adatom and the surface layer of the Am slab were relaxed simultaneously and the system where only the adatom was relaxed with the Am slab being fixed, was found to be of the order of 0.03 eV with no change in site preferences as far as chemisorption was concerned. Thus we expect that surface relaxation effects during adsorption will not be significant in our molecular adsorption study also and will not alter our results qualitatively, if not quantitatively. Also, our recent studies on adsorption on the δ-Pu surface [30] indicated that spin-orbit coupling has negligible effect on adsorption geometry but the binding was slightly stronger with the chemisorption energies increasing by 0.05 to 0.3 eV. Though we have not verified this explicitly for the dhcp Am (0001) surface, we expect the same trend to hold true here. Hence in the current calculations, the geometry was optimized at the NSOC level and the final geometry was used for the SOC calculation to study the effects of spin-orbit coupling on the adsorption energies.

## 3 Results and discussions

Table 1 lists the adsorption energies and the associated geometrical parameters of the oxygen molecule adsorbed on the (0001) surface of dhcp-Am. The differences between the NSOC and SOC chemisorption energies at each adsorption site, given by $\Delta E_c = E_c(SOC) - E_c(NSOC)$, are also listed. We first discuss the Vert approach, where the oxygen molecule approaches the surface at the three different adsorption sites in the vertical molecular orientation. Fig. 3 has the optimized $O_2$ chemisorbed geometries of the americium surface for the Vert approach at the three different adsorption sites. As listed in Table 1 for the Vert approach, the distances from the americium surface ($R_d$) are 2.117 Å, 1.323 Å, and 0.529 Å and the equilibrium O – O bond lengths ($R_O$) are 1.267 Å, 1.418 Å and 1.821 Å for the three adsorption sites t1, b2 and h3, respectively. The most stable site is the three-fold hollow h3 site (3.315 eV for the NSOC case, 3.668 eV for SOC case), followed by the two-fold b2 site (2.638 eV for the NSOC case, 2.811 eV for SOC case), with the least favorable site being the one-fold t1 (1.979 eV for the NSOC case, 2.068 eV for SOC case). The distance between the americium surface and $O_2$ molecule clearly shows that at the least stable t1 site, the admolecule is furthest away from the surface (2.117 Å) followed by the next stable b2 site (1.323 Å), with the distance being the smallest (0.529 Å) for the most stable h3 site.

It follows from the above discussions that increasing stability at both the NSOC and SOC cases implies decreasing vertical distance of the $O_2$ molecule from the surface layer. Also increasing stability implies increasing admolecule coordination number at both theoretical levels, that is, the $O_2$ molecule prefers to bind at the maximally coordinated three-fold hollow h3 site. The O – O bond lengths for the Vert approach at the three adsorption sites shows that at none of the adsorption sites the $O_2$ molecule tends to dissociate Even though the $O_2$ molecule is stretched maximally at the most favorable h3 site (1.821 Å), it cannot be considered as dissociated species.  All chemisorption energies for the Vert approach indicate that binding is slightly stronger with the inclusion of SOC compared to the NSOC case. As mentioned above, the SOC-NSOC chemisorption energy differences $\Delta E_c$ are also listed in Table 1; $\Delta E_c$ is minimum at the least stable t1 site (0.089 eV), followed by the next stable b2 adsorption site (0.173 eV), with the most stable h3 adsorption site having a $\Delta E_c$ of 0.353 eV.

Next we discuss adsorption corresponding to the Hor1 approach, where the $O_2$ molecular orientation is parallel to a lattice vector. In this case, the atoms of the oxygen molecule are at the same distance from the americium surface and $R_d$ is measured from the center of mass of the $O_2$ molecule to the surface. Fig. 4 shows the optimized $O_2$ chemisorbed geometries on the americium surface for the Hor1 approach at the three different adsorption sites. It is known [39-41] that the probability of dissociation of gas molecules on metal surfaces is higher when the molecules are oriented horizontally/parallel with respect to the surface as compared to the case where the molecules are oriented vertically/perpendicularly. This was also found to be true for our Am-$O_2$ system, where the molecule completely dissociates for the Hor1 approach as clearly shown in Fig. 4. Each subfigure is labeled A $\rightarrow$ B + C, where A is the initial adsorption site (the center of mass of the $O_2$ was initially placed at this site) and B and C are the final adsorption sites (each atom of the dissociated molecule sits at this site). Throughout this manuscript, we will use this notation to describe all dissociated configurations. The three dissociated configurations corresponding to the three initial adsorption sites t1, b2 and h3 are: (a) t1 $\rightarrow$ b2 + b2; (b) b2 $\rightarrow$ t1 + t1; (c) h3 $\rightarrow$ b1 + b1, respectively. In Fig. 4(c), the site b1 in transition h3 $\rightarrow$ b1 + b1, is the bridge site derived from the bond between an atom on the surface layer and an atom on the sub-surface layer and as such is one-fold coordinated whereas the site b2 is a bridge site between two surface layer atoms, making it two-fold coordinated.

As listed in Table I for the Hor1 approach, the distances of each O atom from the americium surface $R_d$ are 1.165 Å, 1.905 Å, and 1.587 Å and the O – O separations $R_O$ are 3.517 Å, 3.620 Å and 3.118 Å for the dissociated processes t1 $\rightarrow$ b2 + b2, b2 $\rightarrow$ t1 + t1 and h3 $\rightarrow$ b1 + b1, respectively. Regarding the chemisorption energies, the most stable configuration is the t1 $\rightarrow$ b2 + b2 dissociation (8.681 eV for the NSOC case, 9.191 eV for SOC case), followed by the h3 $\rightarrow$ b1 + b1 dissociation (6.564 eV for the NSOC case, 7.057 eV for SOC case), with the least favorable site being b2 $\rightarrow$ t1 + t1 (5.472 eV for the NSOC case, 5.861 eV for SOC case). Similar to the trend for the Vert approach, increasing stability at both the NSOC and SOC cases implies decreasing vertical distance of the $O_2$ molecule from the surface layer. Analogous to case of Vert approach, the chemisorption energies for the Hor1 approach indicates that binding is slightly stronger with the inclusion of SOC compared to the NSOC case. The SOC-NSOC chemisorption energy differences $\Delta E_c$ are also listed in Table 1; $\Delta E_c$ is minimum at the least stable b2 → t1 + t1 dissociation (0.389 eV) closely followed by the next stable h3 → b1 + b1 dissociation (0.493 eV), with the most stable t1 → b2 + b2 dissociation having the largest difference of $\Delta E_c$ = 0.510 eV.

For the Hor2 approach, where the $O_2$ molecular approach is horizontal but perpendicular to a lattice vector, the adsorption process is dissociative. The final adsorption configurations are shown in Fig. 5. Again $R_d$ is measured from the center of mass of the molecule to the surface. As labeled in Fig. 5, the following dissociative configurations were observed: (a) t1 → h3 + h3; (b) b2 → h3 + f3; (c) h3 → t1 + b2.  It should be noted that in Fig. 5(b) the site f3 corresponds to the three-fold hollow fcc site. The f3 site was not considered as an initial adsorption site since our previous calculations [30,42] have clearly indicated that chemisorption energies at the h3 and f3 sites are nearly degenerate and the geometry (specifically the distance from the surface) is almost identical. This is mainly due to the fact that the h3 and f3 sites have the same local environment (both are three-fold coordinated) except that their relative positions on the surface are different. Thus one hollow site h3 was deemed sufficient for this work. Also, we note that a proper inclusion of the f3 site requires nine layers with 36 Am atoms and an all electron calculation would be computationally prohibitive without generating physically meaningful results.

As listed in Table I, the distances from the americium surface $R_d$ are 0.953 Å, 0.636 Å, and 1.589 Å and the O − O bond lengths ($R_O$) are 3.731 Å, 2.183 Å and 2.569 Å for the three dissociation processes t1 $\rightarrow$ h3 + h3, b2 $\rightarrow$ h3 + f3, and h3 $\rightarrow$ t1 + b2 respectively. The most energetically stable configuration corresponded to the t1 $\rightarrow$ h3 + h3 dissociation (9.395 eV for the NSOC case, 9.886 eV for SOC case), followed by the dissociation b2 $\rightarrow$ h3 + f3 (8.972 eV for the NSOC case, 9.456 eV for SOC case), with the least energetically favorable configuration being the dissociation h3 $\rightarrow$ t1 + b2 (5.615 eV for the NSOC case, 6.084 eV for SOC case). For the most favorable configuration for Hor2 approach, t1 $\rightarrow$ h3 + h3, which is also the most favorable configuration among all the three approaches, we also studied the reaction pathway for the dissociation of oxygen molecule to the most stable hollow sites and no existence of any energy barrier was found. Unlike the trends for the Vert and Hor1 approach, increasing stability at both the NSOC and SOC cases does not necessarily imply decreasing vertical distance of the $O_2$ molecule from the surface layer. However, as observed for the Vert and Hor1 approach, the chemisorption energies for the Hor2 approach indicate that binding is slightly stronger with the inclusion of SOC compared to the NSOC case. The SOC-NSOC chemisorption energy differences $\Delta E_c$ are also listed in Table 1; $\Delta E_c$ is minimum at the least stable h3 $\rightarrow$ t1 + b2 configuration (0.469 eV) closely followed by the next stable b2 $\rightarrow$ h3 + f3 configuration (0.484 eV), with the most stable t1 → h3 + h3 configuration having an SOC-NSOC $\Delta E_c$ = 0.491 eV.

In Table 2, the adsorbate-induced work function changes with respect to the clean metal surface, given by $\Delta\Phi = \Phi^{admolecule/Am} - \Phi^{Am}$, are listed at the NSOC and SOC cases for each adsorbate and each adsorption site. For Vert approach high chemisorption energies generally correspond to high work function shifts. In fact, the changes in the work functions are largest at the most preferred h3 site and lowest at the least preferred adsorption site t1. This though is not true for the Hor1 and Hor2 approaches. For Hor1 and Hor2 approaches, we find that dissociative configurations with high chemisorption energies generally correspond to low work function shifts. For Hor1 approach, the changes in the work functions are largest at the least preferred b2→ t1+t1 configuration and lowest at the most stable t1→ b2+b2 configuration. For Hor2 approach, the change in the work functions is similar to the Hor1 approach, with the largest corresponding to the least preferred h3→ t1+ b2 configuration and lowest change corresponding to the most preferred t1 → h3 + h3 configuration.

The adsorbate-induced work function shifts can be understood in terms of the surface dipoles arising due to the displacement of electron density from the substrate towards the adsorbates since the electronegativity of O is much larger than that of Am. The surface dipole moment µ (in Debye) and the work function shift $\Delta\Phi$ (in eV) are linearly related by the Helmholtz equation $\Delta\Phi = 12\Pi\Theta\mu/A$, where A is the area in $Å^2$ per (1×1) surface unit cell and $\Theta$ is the adsorbate coverage in monolayers. For each of the approaches, Vert, Hor1 and Hor2, and each adsorption configuration, the calculated work function shifts at the NSOC case are found to be slightly larger than that at the SOC case.

In Table 3, the magnitudes and alignments of the site projected spin magnetic moments for each Am atom on the *surface* atomic layer are reported for the clean and adsorbate-covered surfaces. Spin moments were quoted only for the Am atoms on the surface layer because the major changes in the moments after chemisorption occurred on that layer and very little or no changes were observed on the subsurface and central layers. This stems primarily from the fact that the oxygen molecule interacts mainly with the Am atoms on the surface layer. Also, the spin moments reported correspond to the SOC calculations. NSOC moments follow a similar qualitative trend and are not reported here. For each adsorption site, the spin moment of the closest neighbor surface layer Am atoms with which the admolecule primarily interacts is indicated in bold fonts in the Table 3.

For the Vert approach, we see reductions of 0.28 $\mu_B$ for the adsorption site t1, 0.20 $\mu_B$ for each of the two atoms for the adsorption site b2 and 0.20 $\mu_B$ for each of the three atoms for the adsorption site h3 in the spin moment of the Am atom. The moments of the remaining Am atoms which are not bonded to the oxygen molecule remain basically unaltered when compared to the clean surface. For the Hor1 approach, we see spin magnetic moment reductions of 0.60 $\mu_B$ and 0.55 $\mu_B$ for the two Am atoms for the t1$\rightarrow$b2+b2 dissociative configuration, 0.68 $\mu_B$ for each of the two Am atoms for the b2$\rightarrow$ t1+t1 dissociative configuration, and 0.62 $\mu_B$ for each of the two Am atoms for the

h3→b1+b1 dissociative configuration. Finally for the Hor2 approach, we see spin magnetic moment reductions of 0.59 $\mu_B$ for one Am atom and 0.26 $\mu_B$ for two Am atoms for the t1→ h3+h3 dissociation; 0.27 $\mu_B$, 0.26 $\mu_B$ and 0.31 $\mu_B$ for the three Am atoms for the b2→ h3+f3 dissociation, and 0.82 $\mu_B$ for one Am atom and 0.21 $\mu_B$ for two Am atoms for the h3→ t1+b2 dissociation. In all cases, the moments in the interstitial region also decreased after chemisorption. The reduction in the Am magnetic moments after chemisorption is attributed to the transfer of charge from Am to O as predicted by the work function increase after chemisorption.

The nature of the APW+lo basis allows us to decompose the electronic charges inside the muffin-tin spheres into contributions from different angular momentum channels. These charges are referred to as partial charges. Comparing the partial charges $Q_B$ of the Am layers and the adsorbates before adsorption to the corresponding partial charges $Q_A$ after adsorption gives us an idea of the nature of the interaction between the adsorbates and substrate. These are reported in Tables 4-6. For Am, only the *surface layer atoms* were considered as no significant changes were observed on the subsurface and central layers. In each table, we have also reported the differential partial charge of the different angular momentum state *l* corresponding to a given atom given by $\Delta Q(l) = Q_A - Q_B$. $\Delta Q(l) > 0$ indicates charge gain inside the muffin tin sphere while $\Delta Q < 0$ indicates otherwise. $\Delta Q(l)$ may be loosely interpreted as a measure of charge transfer. In Table 4, $Q_A$, $Q_B$, and $\Delta Q(l)$ for $O_2$ molecule with Vert approach adsorbed at the t1, b2, and h3 adsorption sites respectively on the dhcp-Am (0001) surface are reported. For the one-fold t1 site, $\Delta Q(2p) = 0.05$ *e* and 0.01 $e$ for the two O atoms, $\Delta Q$ (6$d$) = 0.09 and $\Delta Q$(5$f$) = -0.14 $e$ for the Am atom, implying Am(6$d$)-Am(5$f$)-O(2$p$) hybridizations. For the two-fold b2 site, $\Delta Q$ (2$p$) = -0.04 $e$ and 0.05 $e$ for the two O atoms, $\Delta Q$(6$d$) = 0.03 $e$ and $\Delta Q$(5$f$) = -0.06 $e$ for each of the two Am atoms, again suggesting the participation of the Am 5$f$ electrons in chemical bonding with O. For the three-fold hollow h3 site, $\Delta Q$ (2$p$) = -0.07 $e$ and 0.08 $e$ for the two O atoms, $\Delta Q$ (6$d$) = 0.04 $e$ and $\Delta Q$ (5$f$) = -0.07 $e$ for each of the three Am atoms, which again suggests contributions of the Am 5$f$ electrons to Am-O bonding. In Table 5, $Q_A$, $Q_B$, and $\Delta Q$ ($l$) for the Hor1 approach corresponding to the t1$\rightarrow$ b2+b2, b2$\rightarrow$ t1+t1, and h3$\rightarrow$ b1+ b1 dissociations are shown. For the t1$\rightarrow$ b2+b2 dissociation, $\Delta Q$ (2$p$) = 0.06 $e$ for the two O atoms, $\Delta Q$(6$d$) = 0.13 $e$ and $\Delta Q$(5$f$) = -0.22 $e$ for each of the two Am atoms, which like the case for Vert approach, implies Am(6$d$)-Am(5$f$)-O(2$p$) interactions. For the b2$\rightarrow$ t1+t1 dissociation, $\Delta Q$ (2$p$) = 0.02 $e$ for the two O atoms, $\Delta Q$ (6$d$) = 0.23 $e$ and $\Delta Q$ (5$f$) = -0.29 $e$ for each of the two Am atoms, suggesting the participation of the Am 5$f$ electrons in Am-O bonding. For the h3$\rightarrow$ b1+ b1 dissociation, $\Delta Q$ (2$p$) = 0.04 $e$ for the two O atoms, $\Delta Q$ (6$d$) = 0.21 $e$ and $\Delta Q$(5$f$) = -0.23 $e$ for each of the two Am atoms, which again suggests contribution of the Am 5$f$ electrons to Am-O chemical bonding.

In Table 6, where the partial charges for the Hor2 approach are reported, we observe for the t1$\rightarrow$h3+h3 dissociation that, $\Delta Q$ (2$p$) = 0.07 $e$ for the two O atoms, $\Delta Q$(6$d$) = 0.02 $e$, 0.02 $e$ and 0.13 $e$ and $\Delta Q$(5$f$) = -0.09 $e$, -0.09 $e$ and -0.22 $e$ for each of the three Am atoms, which like the previous cases for Vert and Hor1 approaches, imply Am(6$d$)-Am(5$f$)-O(2$p$) interactions. For the b2$\rightarrow$

h3+f3 dissociation, $\Delta Q$ ($2p$) = 0.11 $e$ for the two O atoms, $\Delta Q(6d)$ = 0.06 $e$, 0.06 $e$ and 0.07 $e$ and $\Delta Q(5f)$ = -0.11 $e$, -0.11 $e$ and -0.09 $e$ for each of the three Am atoms while for the h3$\rightarrow$ t1+b2 dissociation, $\Delta Q(2p)$ = 0.00 $e$ and 0.06 $e$ for the two O atoms, $\Delta Q(6d)$ = -0.03 $e$ and 0.22 $e$ and $\Delta Q(5f)$ = 0.04 $e$ and -0.24 $e$ for each of the two Am atoms. Overall, the partial charge analyses tend to suggest that *some* of the Am $5f$ electrons participate in chemical bonding. The question of whether all the *5f* electrons are localized or not cannot, of course, be proven from the present analysis. We wish to stress that the partial charges are confined inside the muffin tin spheres and do not give any information of the interactions between the atoms in the interstitial region. Information which includes the electronic charges in interstitial region can be obtained from the difference charge density distributions.

The nature of the bonds between the adsorbate and the Am atoms on the surface has been investigated by computing the difference charge density $\Delta n(r)$ defined as follows:

$$\Delta n(r) = n(O_2 + Am) - n(Am) - n(O_2),$$

where $n(O_2+Am)$ is the total electron charge density of the Am slab with $O_2$ admolecule, $n(Am)$ is the total charge density of the bare Am slab, and $n(O_2)$ is the total charge density of the admolecule. In computing $n(O_2)$ and $n(Am)$, the admolecule $O_2$ and Am atoms are kept fixed at exactly the same positions as they were in the chemisorbed systems. All charge densities reported here were computed in the plane passing through the admolecule and one or two surface Am atoms using the Xcrysden utility [43]. We have reported the difference charge density plots for the most favorable chemisorption site for each of the approaches, Vert, Hor1 and Hor2 in Fig. 6. For the Vert approach, h3 was the most favorable adsorption site and Δn was computed in the plane passing through the two O atoms and the Am atom directly bonded to the O atom facing the surface. For the Hor1 approach, the most stable t1→b2+b2 dissociated configuration is depicted. In this case Δn was computed in the plane passing through the two dissociated O atoms and nearest neighbor Am atoms. Similarly for the Hor2 approach, Δn for the most stable t1→h3+h3 dissociated configuration is shown. Again the plane passes through the two dissociated O atoms and neighboring Am atoms. In Fig. 6(a) we clearly see charge accumulation around the oxygen atom labeled O2 and a strong polarization towards the oxygen atom labeled O1. Also, significant charge loss around the Am atom bonded to the atom O2 implies that the Am-O bond has a strong ionic character, which is expected as the oxygen atom is more electronegative than Am. For the Hor1 approach corresponding to t1→b2+b2 in Fig. 6(b) and Hor2 approach corresponding to t1→ h3+h3 in Fig. 6(c), the Am-O bonds are again largely ionic in character as significant charge accumulation around the O atoms can be observed. The difference charge density plots are fairly consistent with the positive changes in the work function as well as the differential partial charges reported in Tables 4-6.

We also computed the local density of states (LDOS), obtained by decomposing the total density of the single particle Kohn-Sham eigenstates into contributions from each angular momentum channel *l* of the constituent atoms inside the muffin tin spheres. Here, we have reported the LDOS for only the SOC computation, the LDOS for NSOC calculations yielding a similar qualitative description. In Fig. 7, the Gaussian-broadened (with a width of 0.05 eV) $f$ and $d$ LDOS curves for each of the layers of the bare dhcp Am (0001) metal slab are shown. Clearly, we see well-defined peaks in the $5f$ electron LDOS in the vicinity of the Fermi level, which have also been observed for bulk dhcp-Am(0001), and tends to indicate *some* $5f$ electron localization [28]. However, this statement should be viewed with caution because of the nature of the underlying theory, namely density functional theory and all $5f$ states are treated as band states. The occupied $5f$ electron states below the Fermi level also support this conclusion, In view of our previous statement about the participation of the $5f$ electrons in chemical bonding, we believe that it is possible that some $5f$ electrons are localized and some are not. This question needs to be pursued in the future. We also note that the peak below the Fermi level centered on a binding energy of 1 eV below the level instead of the 2.8 eV observed in X-ray and ultraviolet photoemission spectra experiments [14,19]. In Fig. 8, we show the LDOS plots for the $O_2$ molecule and the surface Am atoms after chemisorption for the Vert approach at the three different adsorption sites t1, b2 and h3. As there are four nonequivalent sites on the surface, we depict the LDOS for only the Am atom(s) directly bonded to the $O_2$ molecule (or O atoms for the cases where the molecule dissociates) in order to assess the changes in DOS upon chemisorption. At the adsorption site t1, we note some modification in the Am $5f$ DOS just below the Fermi level in comparison to the $5f$ DOS of the bare surface. More specifically, we observe some reduction in the 5f peak around at 0.5 eV below the Fermi level, implying the participation of *some* 5f electrons participate in chemical bonding.

We also observe an overlap of the Am 5f and 6d bands with the O 2p band in the -2 eV to 0 eV range, implying Am(5*f*)-Am(6*d*)-O(2*p*) hybridizations. We believe that the Am 5f peak located at 1 eV below the Fermi level corresponds to bonding states, but it is not possible, at the level of theory used here, to quantify the amount of 5f electrons actually participating in bonding. The LDOS distributions for the b2 and h3 sites show some modifications in the 5*f* DOS below the Fermi level in comparison to the surface layer LDOS of the bare slab. In particular the splitting of the 5*f* band (about 1 eV below the Fermi level) broadens, implying that some of the 5*f* electrons participate in bonding. Similar to the t1 site, some Am(5f)-Am(6*d*)-O(2*p*) admixture are clearly evident in both cases.

In Fig. 9, we show the LDOS plots for the O atoms and the surface Am atoms after chemisorption for the Hor1 approach corresponding each of the three dissociated configurations. For the dissociation process t1→ b2+b2, we note some broadening of the Am 5*f* DOS just below the Fermi level in comparison to the bare surface layer 5*f* DOS. We also observe some Am(5*f*)-Am(6*d*)-O(2*p*) hybridizations in the -5 eV to -3 eV range, implying that both the Am 5*f* and 6*d* orbitals contribute to bonding. The LDOS distributions for the b2→ t1+t1 and h3→ b1+ b1 dissociations show a slight reduction in the 5*f* DOS below the Fermi level, with the O 2*p* bonding state pushed to slightly higher binding energies (when compared to the DOS for t1→ b2+b2), which naturally suggests slightly weaker binding as observed in the chemisorption energies.

In Fig. 10, we show the LDOS plots for the dissociated molecule and the surface Am atoms after chemisorption for the Hor2 approach corresponding to the t1→ h3+h3, b2→ h3+f3, and h3→ t1+ b2. In all cases, we observe a broadening in the 5*f* band peaks below the Fermi level when compared to the bare surface. Furthermore, some O(2*p*) and Am(6*d*, 5*f*) hybridizations are evident. We hasten to point out that overall, the Am 5*f* states for all the Vert, Hor1, and Hor2 approaches show signatures of *slight* delocalization.

## 4. Conclusions

In summary, we have used the generalized gradient approximation to density functional theory with the full potential LAPW+lo method to study chemisorption of oxygen molecule on the (0001) surface of dhcp Am at two theoretical levels; one with no spin-orbit coupling (NSOC) and the other with spin-orbit coupling (SOC). The results at the two cases do not vary from each other significantly. Dissociative adsorption of oxygen molecule is favored over molecular adsorption. For $O_2$ adsorption, the one-fold t1 site with Hor2 approach was found to be the most stable where upon adsorption of the $O_2$ molecule dissociated and two O atoms sit on two different three-fold h3 sites. The inclusion of spin-orbit coupling lowers the chemisorption energies by 0.089-0.493 eV. Work functions increased in all cases compared to the clean Am surface, with the lowest shift corresponding to the least coordinated t1 site and largest shifts corresponding to the maximally coordinated hollow h3 site for the Vert and Hor2 approaches, while for Hor1 approach, this was not the case. Upon adsorption, the net spin magnetic moment of the chemisorbed system decreases in each case compared to the bare surface. The partial charge analyses illustrate that *some* of the Am 5*f* electrons participate in chemical bonding. Difference charge density distributions clearly show that bonds between the surface Am atoms and the oxygen molecule at each site is largely ionic in character. A study of the local density of states showed Am(6*d*)-Am(5*f*)-O(2*p*) hybridizations after chemisorption. Overall, the Am 5*f* DOS below the Fermi Level become slightly delocalized after chemisorption.

Finally, we expect chemisorption energies reported here to be higher than sthe chemisorption energies obtained from a theory in which all the Am *5f* electrons are treated as localized states. The resulting difference in the chemisorption energies can be used to gauge the degree of participation of the 5*f* electrons in chemical bonding. Thus the chemisorption energies reported here can be used as upper bounds for the binding energies of oxygen on dhcp americium surfaces.

## Acknowledgments

This work is supported by the Chemical Sciences, Geosciences and Biosciences Division, Office of Basic Energy Sciences, Office of Science, U. S. Department of Energy (Grant No. DE-FG02-03ER15409) and the Welch Foundation, Houston, Texas (Grant No. Y-1525). In addition to the supercomputing facilities at the University of Texas at Arlington, this research also used resources of the National Energy Research Scientific Computing Center, which is supported by the Office of Science of the U.S. Department of Energy under Contract No. DE-AC02-05CH11231. The authors also acknowledge the Texas Advanced Computing Center (TACC) at the University of Texas at Austin (http://www.tacc.utexas.edu) for providing computational resources.

**References**

[1] http://nobelprize.org/nobel_prizes/chemistry/laureates/2007/chemadv07.pdf

[2] J. J. Katz, G. T. Seaborg, and L. R. Morss, *The Chemistry of the Actinide Elements* (Chapman and Hall, 1986); L. R. Morss and J. Fuger, Eds. *Transuranium Elements: A Half Century* (American Chemical Society, Washington, D. C. 1992).

[3] L. R. Morss, N. M. Edelstein, and J. Fuger, Eds; J. J. Katz, Hon. Ed. *Chemistry of the Actinide and Transactinide Elements* (Springer, New York, 2006).

[4] *Challenges in Plutonium Science,* Vol. I and II, Los Alamos Science, **26** (2000).

[5] R. Haire, S. Heathman, M. Idiri, T. Le Bihan, and A. Lindbaum, Nuclear Materials Technology/Los Alamos National Laboratory, 3rd/4th quarter 2003, p. 23.

[6] *Fifty Years with Transuranium Elements*, Proceedings of the Robert A. Welch Foundation, October 22-23, 1990, Houston, Texas.

[7] J. L. Sarrao, A. J. Schwartz, M. R. Antonio, P. C. Burns, R. G. Haire, and H. Nitsche, Eds. *Actinides 2005-Basic Science, Applications, and Technology,* Proceedings of the Materials Research Society, **893** (2005).

[8] K. J. M. Blobaum, E.A. Chandler, L. Havela, M. B. Maple, M. P. Neu, Eds. *Actinides 2006-Basic Science, Applications, and Technology,* Proceedings of the Materials Research Society, **986** (2006).

[9] *The Elements beyond Uranium,* Glenn T. Seaborg and Walter D. Loveland, p. 17, (John Wiley & Sons, Inc. 1990).

[10] S. Y. Savrasov, K. Haule, and G. Kotliar, Phys. Rev. Lett. **96** (2006) 036404.

[11] S. Heathman, R. G. Haire, T. Le Bihan, A. Lindbaum, K. Litfin, Y. Méresse, and H. Libotte, Phys. Rev. Lett. **85** (2000) 2961.

[12] G. H. Lander and J. Fuger, Endeavour, **13** (1989) 8.

[13] H. L. Skriver, O. K. Andersen, and B. Johansson, Phys. Rev. Lett. **41** (1978) 42.

[14] J. R. Naegele, L. Manes, J. C. Spirlet, and W. Müller, Phys. Rev. Lett. **52** (1984) 1834.

[15] A. Lindbaum, S. Heathman, K. Litfin and Y. Méresse, Phys. Rev. B. **63** (2001) 214101.

[16] M. Pénicaud, J. Phys. Cond. Matt. **14** (2002) 3575; *ibid*, **17** (2005) 257.

[17] P. Sõderlind, R. Ahuja, O. Eriksson, B. Johansson, and J. M. Wills, Phys. Rev. B. **61** (2000) 8119; P. Sõderlind and A. Landa, *ibid*, **72** (2005) 024109.

[18] P. G. Huray, S. E. Nave, and R. G. Haire, J. Less-Com. Met. **93** (1983) 293.

[19] T. Gouder, P. M. Oppeneer, F. Huber, F. Wastin, and J. Rebizant, Phys. Rev. B **72** (2005) 115122.

[20] L. E. Cox, J. W. Ward, and R. G. Haire, Phys. Rev. B **45** (1992) 13239.

[21] O. Eriksson and J. M. Wills, Phys. Rev. B **45** (1992) 3198.

[22] A. L. Kutepov, and S. G. Kutepova, J. Magn. Magn. Mat. **272-276** (2004) e329.

[23] A. Shick, L. Havela, J. Kolorenc, V. Drchal, T. Gouder, and P. M. Oppeneer, Phys. Rev. B **73** (2006) 104415.

[24] S. Y. Savrasov, G. Kotliar, and E. Abrahams, Nature **410** (2001) 793; G. Kotliar and D. Vollhardt, Phys. Today **57** (2004) 53; X. Dai, S. Y. Savrasov, G. Kotliar, A. Migliori, H. Ledbetter, and E. Abrahams, Science **300** (2003) 953.

[25] B. Johansson and A. Rosengren, Phys. Rev. B **11** (1975) 2836.

[26] J. L. Smith and R. G. Haire, Science **200** (1978) 535.

[27] J. C. Griveau, J. Rebizant, G. H. Lander, and G. Kotliar, Phys. Rev. Lett. **94** (2005) 097002.

[28] D. Gao and A. K. Ray, Eur. Phys. J. B, **50** (2006) 497; MRS Fall 2005 Symp. Proc. **893** (2006) 39; Surf. Sci., **600** (2006) 4941; Eur. Phys. J. B **55** (2007) 13; Phys. Rev. B **77** (2008) 035123.

[29] P. P. Dholabhai, R. Atta-Fynn and A. K. Ray, **61** (2008) 261.

[30] R. Atta-Fynn and A. K. Ray, Phys. Rev. B **75** (2007) 195112 and references therein; Physica B **400** (2007) 307.

[31] P. Hohenberg and W. Kohn, Phys. Rev. **136** (1964) B864; W. Kohn and L. J. Sham, Phys. Rev. **140** (1965) A1133.

[32] J. P. Perdew, K. Burke, and M. Ernzerhof, Phys. Rev. Lett. **77** (1996) 3865.

[33] P. Blaha, K. Schwarz, G. K. H. Madsen, D. Kvasnicka, and J. Luitz, *WIEN2k, An Augmented Plane Wave Plus Local Orbitals Program for Calculating Crystal properties* (Vienna University of Technology, Austria, 2001).

[34] D. D. Koelling and B. N. Harmon, J. Phys. C **10** (1977) 3107.

[35] J. Kunes, P. Novak, R. Schmid, P. Blaha, and K. Schwarz, Phys. Rev. B **64** (2001) 153102.

[36] F. D. Murnaghan, Proc. Natl. Acad. Sci. USA **30** (1944) 244.

[37] R. W. G. Wyckoff, *Crystal Structure*s Volume 1 (Wiley, New York, 1963).

[38] F. Wagner, Th. Laloyaux, and M. Scheffler, Phys. Rev. B **57** (1998) 2102; J. L. F. Da Silva, C. Stampfl, and M. Scheffler, Surf. Sci. **600** (2006) 703.

[39] B. G. Briner, M. Doering, H.-P. Rust, and A. M. Bradshaw, Phys. Rev. Lett. **78** (1997) 1516.

[40] S. Y. Liem, J. H. R. Clarke, and G. Kresse, Comp. Mat. Sci. **17** (2000) 133.

[41] G. Katz, R. Kosloff, and Y. Zeiri, J. Chem. Phys **120** (2004) 3931.

[42] M. N. Huda and A. K. Ray. Int. J. Quant. Chem. **105** (2005) 280.

[43] A. Kokalj, J. Mol. Graph. Mod. **17**(1999)176; code available from http://www.xcrysden.org

Table 1: Adsorbate geometry and adsorption energies at the different adsorption sites. $E_c$ is the chemisorption energy, $R_d$ is the perpendicular distance of the oxygen molecule from the surface, $R_O$ is the O − O separation and D is Am-O nearest neighbor distance. The difference between the SOC and NSOC chemisorption energies is given by $\Delta E_c = E_c(SOC) - E_c(NSOC)$. For the dissociative configurations corresponding to the Hor1 and Hor2 approach, the site labeling A→ B+C refers to an initial adsorption site A and final dissociation adsorption sites B and C.

| $O_2$ Approach | Site | $E_c$ (eV) (NSOC) | $E_c$ (eV) (SOC) | $R_d$(Å) | $R_O$(Å) | D (Å) | $\Delta E_c$ (eV) |
|---|---|---|---|---|---|---|---|
| Vert | t1 | 1.979 | 2.068 | 2.117 | 1.267 | 2.117 | 0.089 |
| | b2 | 2.638 | 2.811 | 1.323 | 1.418 | 2.213 | 0.173 |
| | h3 | 3.315 | 3.668 | 0.529 | 1.821 | 2.115 | 0.353 |
| Hor1 | t1→ b2+b2 | 8.681 | 9.191 | 1.165 | 3.517 | 2.109 | 0.510 |
| | b2→ t1+t1 | 5.472 | 5.861 | 1.905 | 3.620 | 1.905 | 0.389 |
| | h3→ b1+ b1 | 6.564 | 7.057 | 1.587 | 3.118 | 1.901 | 0.493 |
| Hor2 | t1→ h3+h3 | 9.395 | 9.886 | 0.953 | 3.731 | 2.094 | 0.491 |
| | b2→ h3+f3 | 8.972 | 9.456 | 0.636 | 2.183 | 2.079 | 0.484 |
| | h3→ t1+ b2 | 5.615 | 6.084 | 1.589 | 2.569 | 1.762 | 0.469 |

Table 2: Change in work function $\Delta\Phi = \Phi^{admolecule/Am} - \Phi^{Am}$ (in eV) for both the NSOC and SOC cases, where $\Phi^{Am}$ is work function of the bare surface and $\Phi^{admolecule/Am}$ is the work function of the surface-with-admolecule. $\Phi^{Am} = 2.906$ eV and 2.989 eV for the NSOC and SOC cases, respectively.

| $O_2$ Approach | Site | $\Delta\Phi$ (eV) NSOC | $\Delta\Phi$ (eV) SOC |
|---|---|---|---|
| Vert | t1 | 1.951 | 1.945 |
| | b2 | 1.983 | 1.952 |
| | h3 | 2.053 | 2.027 |
| Hor1 | t1→ b2+b2 | 1.103 | 1.005 |
| | b2→ t1+t1 | 3.086 | 3.009 |
| | h3→ b1+ b1 | 2.336 | 2.211 |
| Hor2 | t1→ h3+h3 | 0.783 | 0.621 |
| | b2→ h3+f3 | 0.884 | 0.719 |
| | h3→ t1+ b2 | 2.294 | 2.187 |

Table 3: Site projected SOC magnetic moments of surface Am atoms. $\mu$ ($\mu_B$) is the site projected spin magnetic moments for the Am atoms at the surface layer of the bare slab and the chemisorbed systems.

|  | Site | $\mu$ ($\mu_B$) |  | Site | $\mu$ ($\mu_B$) |  | Site | $\mu$ ($\mu_B$) |
|---|---|---|---|---|---|---|---|---|
| Bare Slab |  | 5.81, 5.81<br>5.81, 5.81 |  |  |  |  |  |  |
| Vert | t1 | **5.53**, 5.80<br>5.80, 5.80 | Hor1 | t1→b2+b2 | **5.21**, **5.26**<br>5.80, 5.80 | Hor2 | t1→ h3+h3 | **5.22**, **5.55**<br>**5.55**, 5.82 |
| | b2 | **5.61**, **5.61**<br>5.80, 5.80 | | b2→ t1+t1 | **5.13**, **5.13**<br>5.80, 5.80 | | b2→ h3+f3 | **5.54**, **5.55**<br>**5.50**, 5.80 |
| | h3 | **5.61**, **5.61**<br>**5.61**, 5.80 | | h3→ b1+b1 | **5.19**, **5.19**<br>5.80, 5.80 | | h3→ t1+b2 | **4.99**, **5.60**<br>**5.60**, 5.80 |

Table 4: Distribution of partial charges of O and Am inside the muffin tin sphere at the **t1, b2 and h3** sites for the **Vert** approach for oxygen molecule. $Q_B$ and $Q_A$ are the partial charges inside the muffin tin spheres before adsorption and after adsorption respectively. $\Delta Q = Q_A - Q_B$ is the difference between partial charges. Results correspond to the SOC case. The surface layer Am atoms bonded to O atoms are given in bold fonts.

| Vert approach | Partial charges in muffin-tin | | | | | | $\Delta Q = Q_A - Q_B$ | | |
|---|---|---|---|---|---|---|---|---|---|
| | Before adsorption $Q_B$ | | | After adsorption $Q_A$ | | | | | |
| | O $p$ | Am $d$ | Am $f$ | O $p$ | Am $d$ | Am $f$ | O $p$ | Am $d$ | Am $f$ |
| t1 | 2.03 | | | 2.08 | | | 0.05 | | |
| | 2.03 | | | 2.04 | | | 0.01 | | |
| | | 0.27 | 5.85 | | 0.27 | 5.88 | | 0.00 | 0.03 |
| | | 0.27 | 5.85 | | 0.27 | 5.88 | | 0.00 | 0.03 |
| | | 0.27 | 5.85 | | **0.36** | **5.71** | | 0.09 | -0.14 |
| | | 0.27 | 5.85 | | 0.27 | 5.88 | | 0.00 | 0.03 |
| b2 | 2.03 | | | 1.99 | | | -0.04 | | |
| | 2.03 | | | 2.08 | | | 0.05 | | |
| | | 0.27 | 5.85 | | **0.30** | **5.79** | | 0.03 | -0.06 |
| | | 0.27 | 5.85 | | **0.30** | **5.79** | | 0.03 | -0.06 |
| | | 0.27 | 5.85 | | 0.26 | 5.88 | | -0.01 | 0.03 |
| | | 0.27 | 5.85 | | 0.26 | 5.88 | | -0.01 | 0.03 |
| h3 | 2.03 | | | 1.96 | | | -0.07 | | |
| | 2.03 | | | 2.11 | | | 0.08 | | |
| | | 0.27 | 5.85 | | **0.31** | **5.78** | | 0.04 | -0.07 |
| | | 0.27 | 5.85 | | **0.31** | **5.78** | | 0.04 | -0.07 |
| | | 0.27 | 5.85 | | **0.31** | **5.78** | | 0.04 | -0.07 |
| | | 0.27 | 5.85 | | 0.26 | 5.87 | | -0.01 | 0.02 |

Table 5: Distribution of partial charges of O and Am inside the muffin tin sphere at the **t1, b2 and h3** sites for the **Hor1** approach for oxygen molecule. Description of parameters and atoms is the same as in Table 4.

| Hor1 approach | Partial charges in muffin-tin | | | | | | $\Delta Q = Q_A - Q_B$ | | |
| | Before adsorption $Q_B$ | | | After adsorption $Q_A$ | | | | | |
| | O $p$ | Am $d$ | Am $f$ | O $p$ | Am $d$ | Am $f$ | O $p$ | Am $d$ | Am $f$ |
|---|---|---|---|---|---|---|---|---|---|
| | 2.03 | | | 2.09 | | | 0.06 | | |
| | 2.03 | | | 2.09 | | | 0.06 | | |
| t1→b2+b2 | | 0.27 | 5.85 | | 0.24 | 5.91 | | -0.03 | 0.06 |
| | | 0.27 | 5.85 | | **0.40** | **5.63** | | 0.13 | -0.22 |
| | | 0.27 | 5.85 | | **0.40** | **5.63** | | 0.13 | -0.22 |
| | | 0.27 | 5.85 | | 0.24 | 5.91 | | -0.03 | 0.06 |
| | 2.03 | | | 2.05 | | | 0.02 | | |
| | 2.03 | | | 2.05 | | | 0.02 | | |
| b2→ t1+t1 | | 0.27 | 5.85 | | **0.50** | **5.56** | | 0.23 | -0.29 |
| | | 0.27 | 5.85 | | **0.50** | **5.56** | | 0.23 | -0.29 |
| | | 0.27 | 5.85 | | 0.24 | 5.90 | | -0.03 | 0.05 |
| | | 0.27 | 5.85 | | 0.24 | 5.90 | | -0.03 | 0.05 |
| | 2.03 | | | 2.07 | | | 0.04 | | |
| | 2.03 | | | 2.07 | | | 0.04 | | |
| h3→ b1+b1 | | 0.27 | 5.85 | | **0.48** | **5.62** | | 0.21 | -0.23 |
| | | 0.27 | 5.85 | | **0.48** | **5.62** | | 0.21 | -0.23 |
| | | 0.27 | 5.85 | | 0.24 | 5.88 | | -0.03 | 0.03 |
| | | 0.27 | 5.85 | | 0.24 | 5.88 | | -0.03 | 0.03 |

Table 6: Distribution of partial charges of O and Am inside the muffin tin sphere at the **t1, b2 and h3** sites for the **Hor2** approach for oxygen molecule. Description of parameters and atoms is the same as in Table 4.

| Hor2 approach | Partial charges in muffin-tin | | | | | | $\Delta Q = Q_A - Q_B$ | | |
| --- | --- | --- | --- | --- | --- | --- | --- | --- | --- |
| | Before adsorption $Q_B$ | | | After adsorption $Q_A$ | | | | | |
| | O $p$ | Am $d$ | Am $f$ | O $p$ | Am $d$ | Am $f$ | O $p$ | Am $d$ | Am $f$ |
| | 2.03 | | | 2.10 | | | 0.07 | | |
| | 2.03 | | | 2.10 | | | 0.07 | | |
| t1→h3+h3 | | 0.27 | 5.85 | | **0.29** | **5.76** | | 0.02 | -0.09 |
| | | 0.27 | 5.85 | | **0.29** | **5.76** | | 0.02 | -0.09 |
| | | 0.27 | 5.85 | | **0.40** | **5.63** | | 0.13 | -0.22 |
| | | 0.27 | 5.85 | | 0.27 | 5.87 | | 0.00 | 0.02 |
| | 2.03 | | | 2.14 | | | 0.11 | | |
| | 2.03 | | | 2.14 | | | 0.11 | | |
| b2→ h3+f3 | | 0.27 | 5.85 | | **0.33** | **5.74** | | 0.06 | -0.11 |
| | | 0.27 | 5.85 | | **0.33** | **5.74** | | 0.06 | -0.11 |
| | | 0.27 | 5.85 | | **0.34** | **5.76** | | 0.07 | -0.09 |
| | | 0.27 | 5.85 | | 0.27 | 5.86 | | 0.00 | 0.01 |
| | 2.03 | | | 2.03 | | | 0.00 | | |
| | 2.03 | | | 2.09 | | | 0.06 | | |
| h3→ t1+b2 | | 0.27 | 5.85 | | 0.27 | 5.83 | | 0.00 | -0.02 |
| | | 0.27 | 5.85 | | 0.27 | 5.83 | | 0.00 | -0.02 |
| | | 0.27 | 5.85 | | **0.24** | **5.89** | | -0.03 | 0.04 |
| | | 0.27 | 5.85 | | **0.59** | **5.61** | | 0.22 | -0.24 |

**(a) Bare (0001) surface of dhcp-Am (side view)**

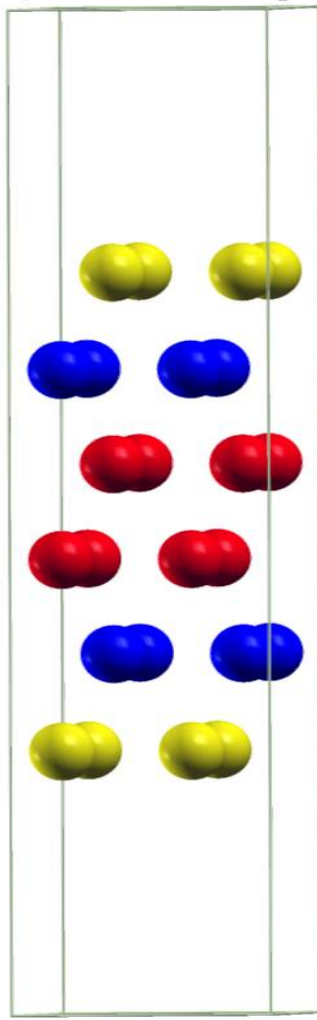

**(b) Bare (0001) surface of dhcp-Am (top view)**

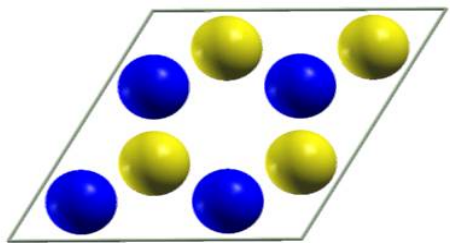

Fig. 1 (Color online) Side view and top view illustrations of six layers of bare (0001) surface of dhcp-Am.

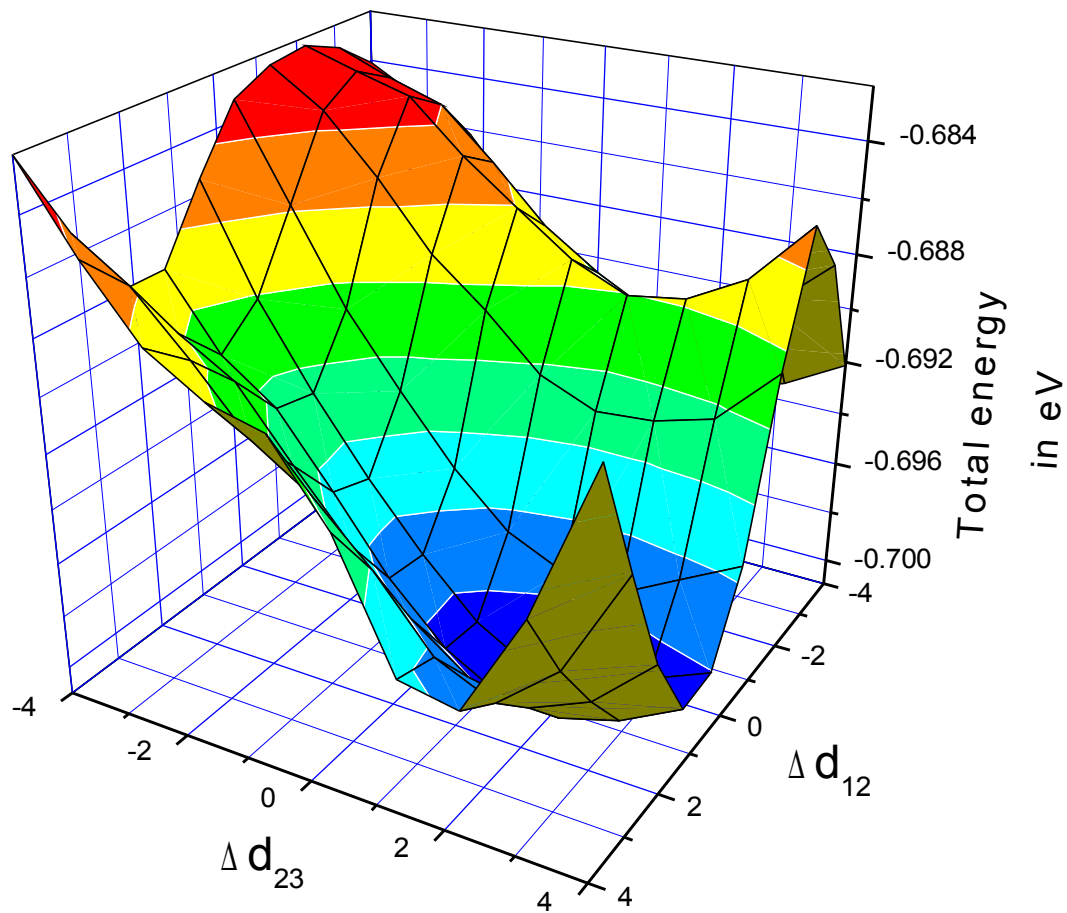

Fig. 2 (Color online) Relaxation of (0001) surface of dhcp-Am. $\Delta d_{12}$ is the percent relaxation of the subsurface layer and $\Delta d_{23}$ is the percent relaxation of the surface layer. Total energy is shifted by a constant factor.

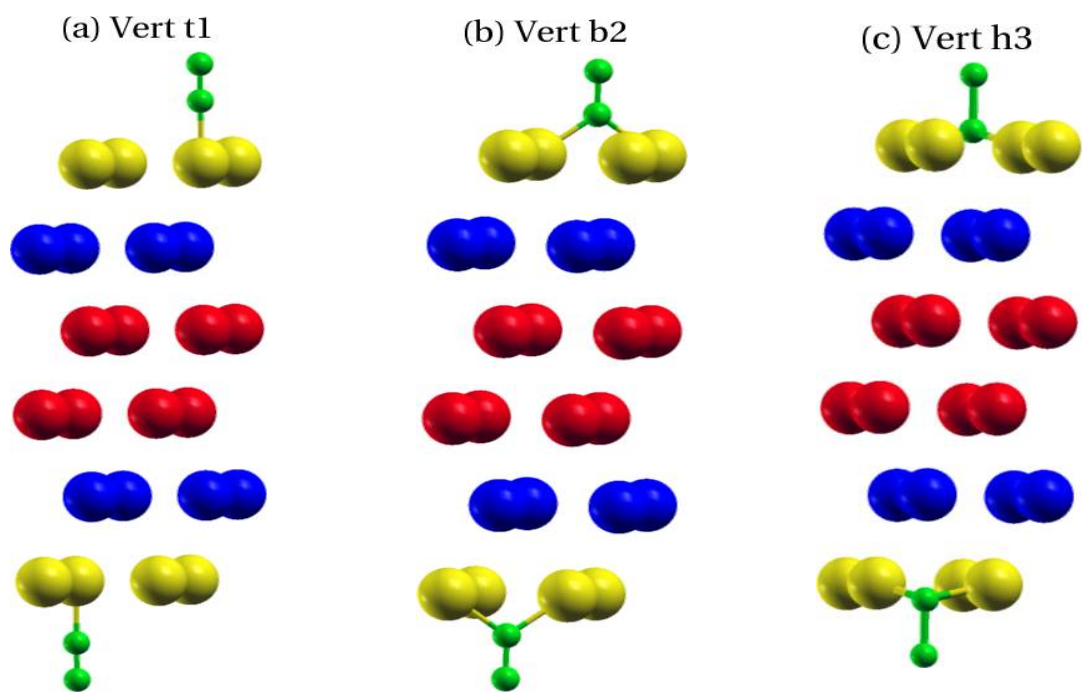

Fig. 3 (Color online) Side view of $O_2$ molecular adsorption on the Am surface at three different adsorption sites for the Vert approach: (a) one-fold top site t1; (b) two-fold bridge site b2; (c) three-fold hollow site h3.

**(a) Hor1 t1 --> b2 + b2**

**(b) Hor1 b2 --> t1 + t1**

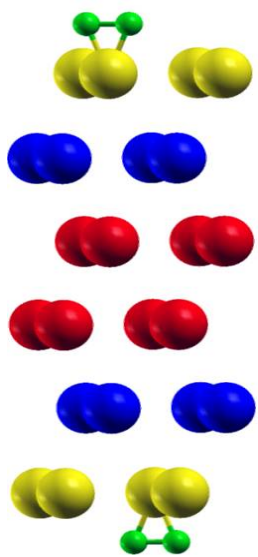

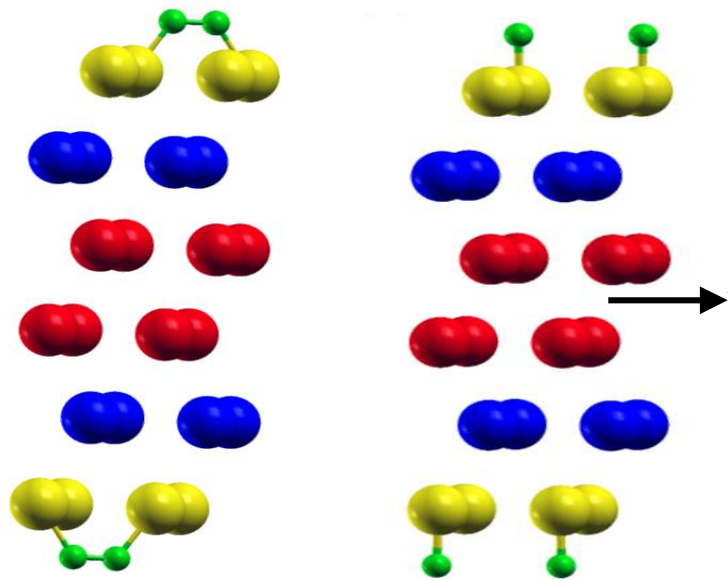

**(c) Hor1 h3 --> b1 + b1**

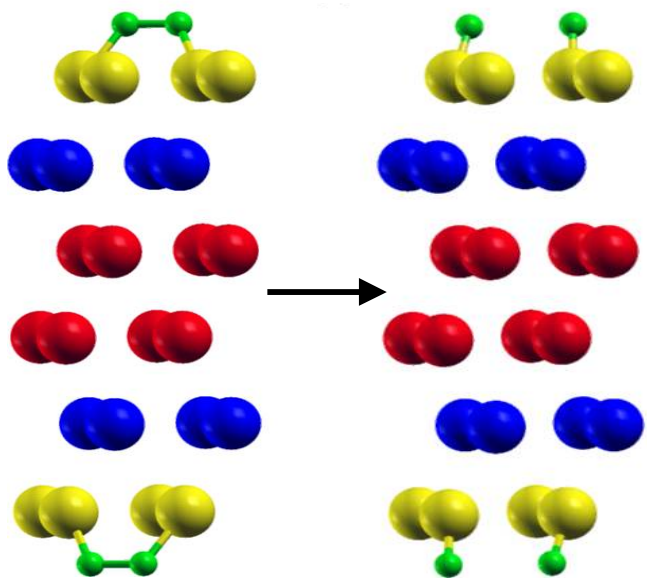

Fig. 4 (Color online) Side view illustrations for the dissociation of $O_2$ molecule on the Am surface for the Hor1 approach: (a) initial site t1, final sites b2+b2; (b) initial site b2, final sites t1+t1; (c) initial site h3, final sites b1+b1.

(a) Hor2 t1 --> h3 + h3

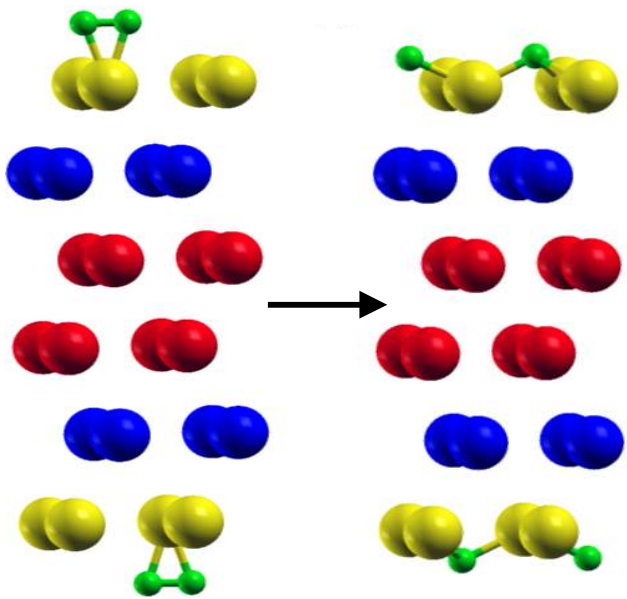

(b) Hor2 b2 --> h3 + f3

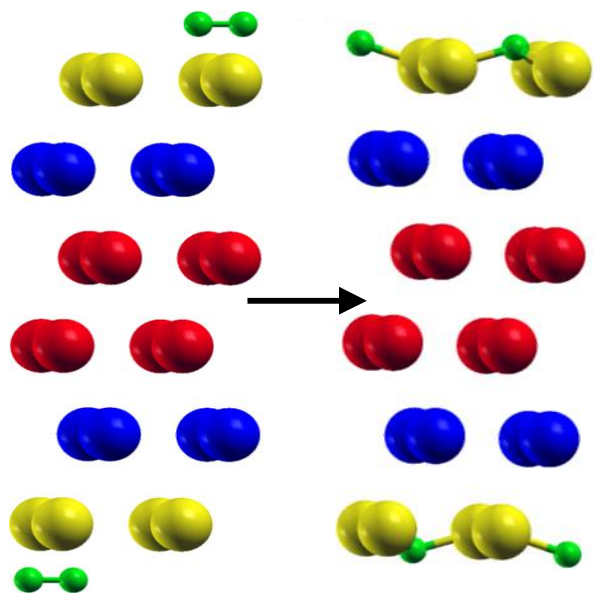

(c) Hor2 h3 --> t1 + b2

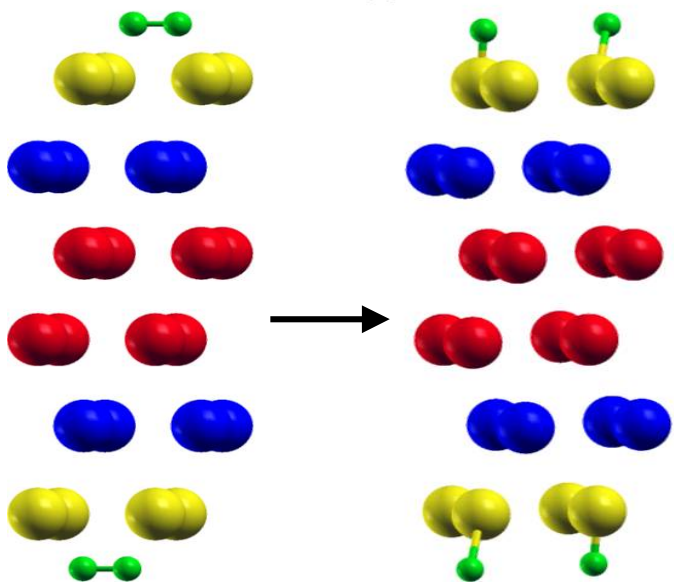

Fig. 5 (Color online) Side view illustrations for the dissociation of $O_2$ molecule on the Am surface for the Hor2 approach: (a) initial site t1, final sites h3+b3; (b) initial site b2, final sites h3+f3; (c) initial site h3, final sites t1+b2.

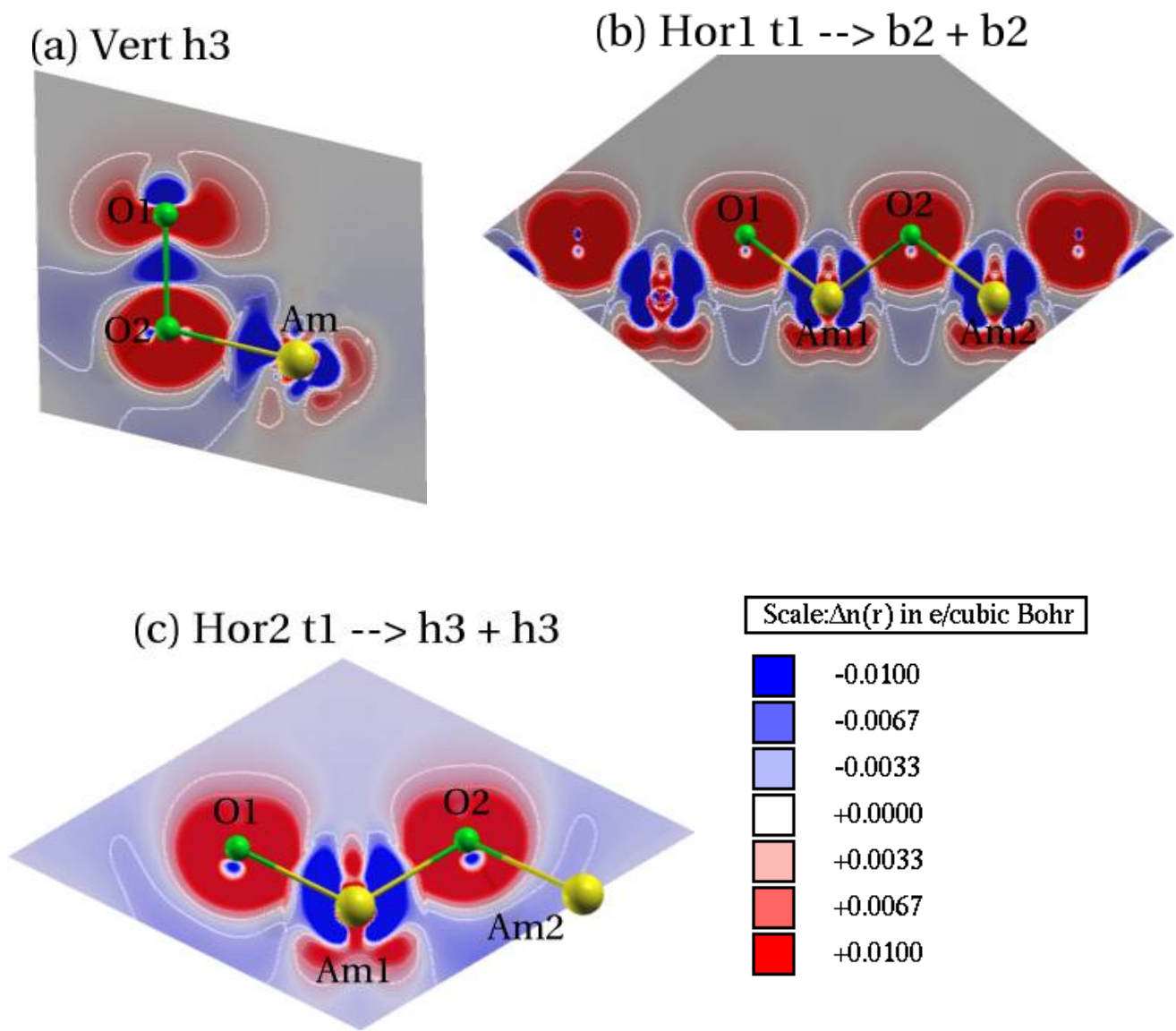

Fig. 6 (Color online) Difference charge density distributions Δn(r) for $O_2$ chemisorbed on the dhcp-Am(0001) surface at the most stable configurations corresponding to the Vert, Hor1 and Hor2 approaches. The scale used is shown at the bottom. Red (positive) denotes regions of charge accumulation and blue (negative) denotes regions of charge loss. Admolecule is colored green and Am atoms are colored gold.

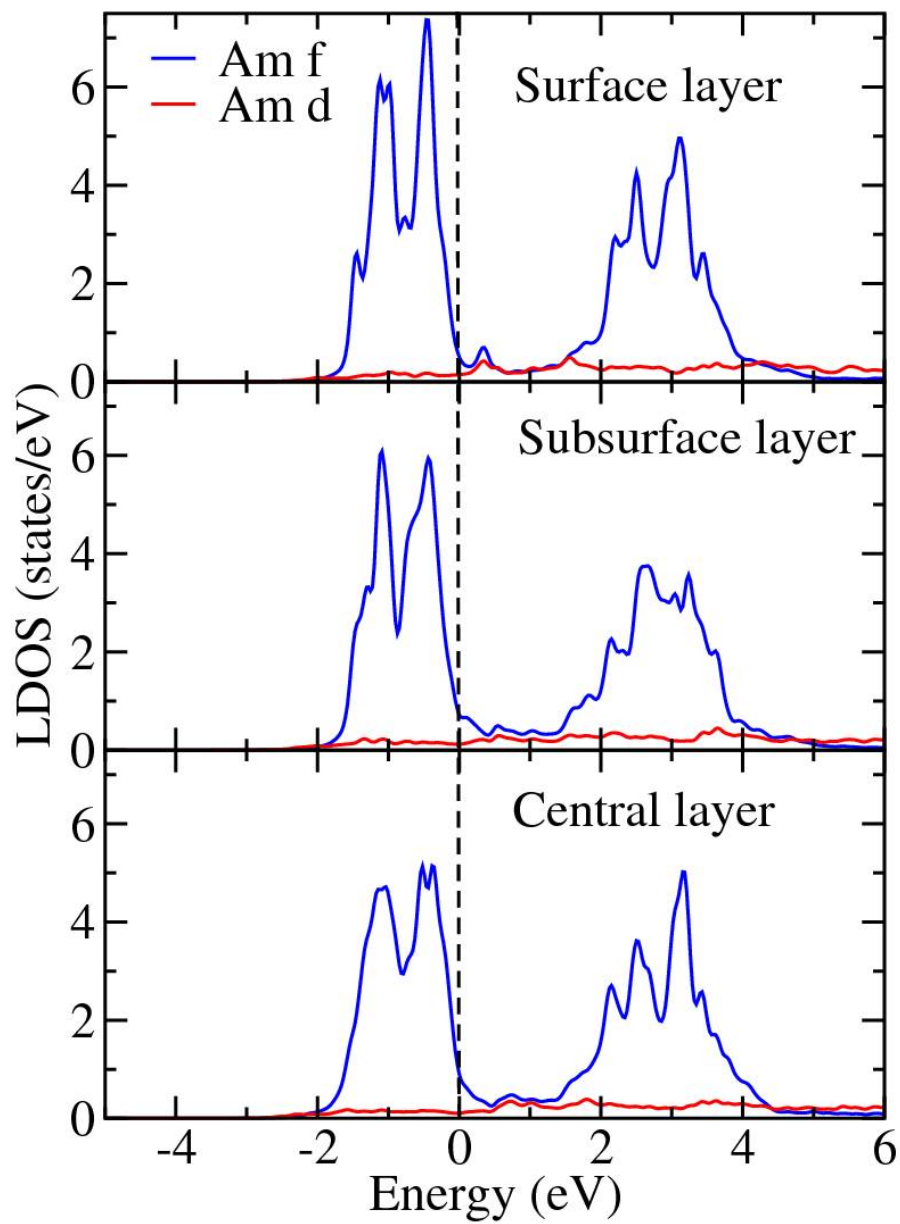

Fig. 7 (Color Online) 6*d* and 5*f* LDOS curves inside the muffin-tins for each layer of the bare dhcp-Am(0001) slab. Vertical line through E=0 is the Fermi level. LDOS correspond to calculations with SOC.

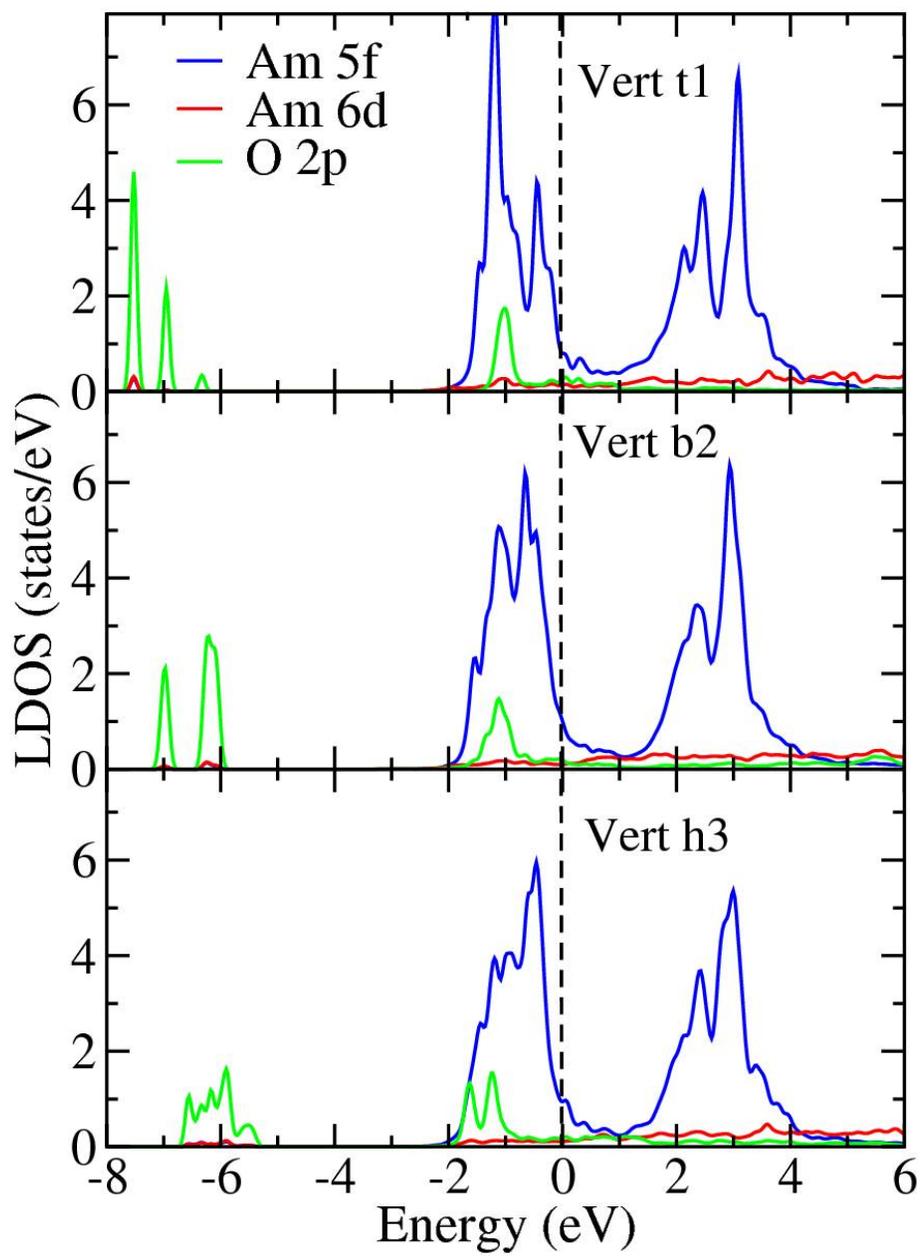

Fig. 8 (Color Online): 6*d* and 5*f* LDOS curves inside the muffin-tins for the Am atoms on the surface layer and 2*p* LDOS curves for $O_2$ admolecule for the Vert approach. Vertical line through E=0 is the Fermi level. LDOS correspond to calculations with SOC.

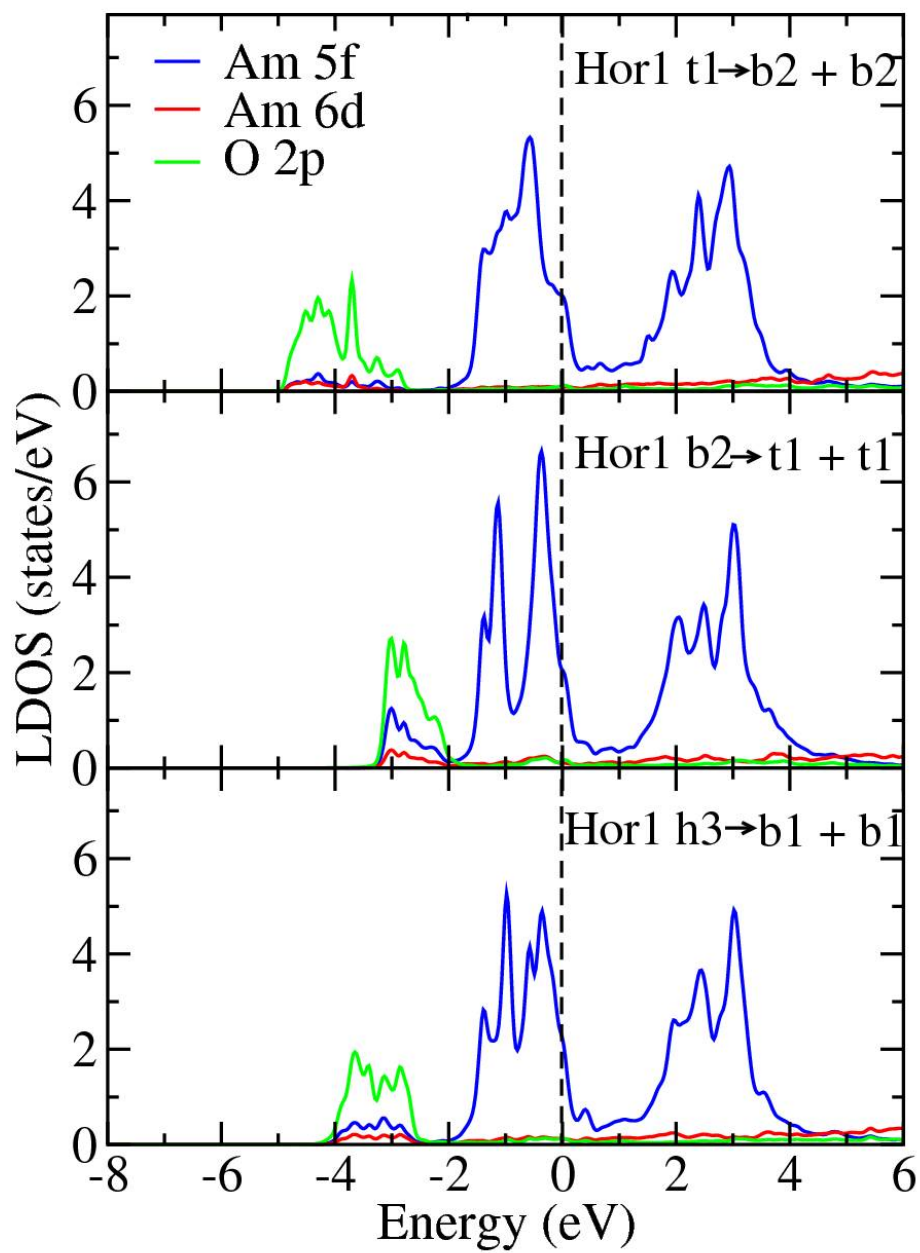

Fig. 9 (Color Online): 6*d* and 5*f* LDOS curves for the Am atoms on the surface layer and 2*p* LDOS curves for O atoms inside the muffin-tins for the Hor1 approach. Vertical line through E=0 is the Fermi level. LDOS correspond to calculations with SOC.

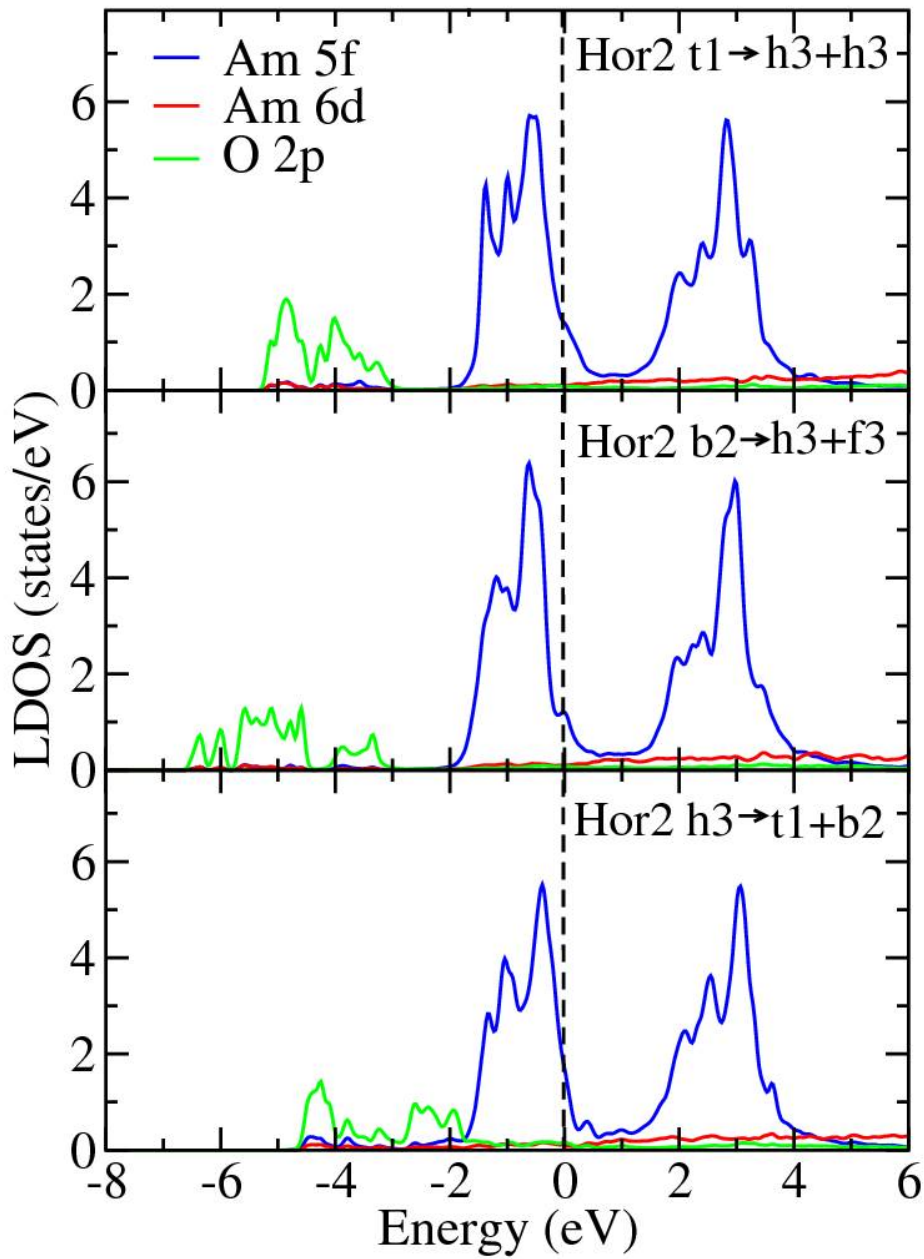

Fig. 10 (Color Online): 6*d* and 5*f* LDOS curves for the Am atoms on the surface layer and 2*p* LDOS curves for O atoms inside the muffin-tins for the Hor2 approach. Vertical line through E=0 is the Fermi level. LDOS correspond to calculations with SOC.